\documentclass[aps,pra,superscriptaddress,twocolumn]{revtex4}
\usepackage{float}
\usepackage{bm}
\usepackage{graphicx,amsmath,latexsym,amssymb}
\usepackage{color}
\usepackage{dcolumn}
\usepackage[utf8x]{inputenc}

\begin{document}

\title[TWEEPR]{ Traveling wave enantio-selective electron paramagnetic resonance}
\author{M. Donaire} 
\email{manuel.donaire@uva.es}
\affiliation{Departamento de F\'isica Te\'orica, At\'omica y \'Optica and IMUVA,  Universidad de Valladolid, Paseo Bel\'en 7, 47011 Valladolid, Spain}
\author{N. Bruyant} 
\affiliation{Laboratoire National des Champs Magn\'{e}tiques Intenses UPR3228 CNRS/EMFL/INSA/UGA/UPS, Toulouse \& Grenoble,France}
\author{G.L.J.A. Rikken} 
\email{geert.rikken@lncmi.cnrs.fr}
\affiliation{Laboratoire National des Champs Magn\'{e}tiques Intenses UPR3228 CNRS/EMFL/INSA/UGA/UPS, Toulouse \& Grenoble,France}

\begin{abstract}
We propose a novel method for enantio-selective electron paramagnetic
resonance spectroscopy based on magneto-chiral anisotropy. We calculate the
strength of this effect and propose a dedicated interferometer setup for its
observation.
\end{abstract}

\date{\today }
\maketitle



\noindent \emph{Introduction}\newline
Electron paramagnetic resonance (EPR) spectroscopy is a powerful technique
to study the local environment and the dynamics of spin-carrying entities,
like transition metal ion complexes and organic radicals \cite{EPR review}.
Also, those systems that do not intrinsically carry a spin can still be
studied by EPR through spin-labelling, i.e., by selectively adding-on a spin
carrying probe \cite{Spin label review}. Many of the systems studied by EPR
are chiral, i.e., they exist in two non-superimposable forms (enantiomers)
that are each other's mirror image, particularly in biochemistry where
enzymes, metalloproteins, membranes, etc., are chiral subjects of intense
EPR activity \cite{Biochem EPR}.\ However, EPR is universally believed to be
blind to chirality. Here we present the paradigm shift that EPR in the
proper configuration is intrinsically sensitive to chirality because of
magneto-chiral anisotropy (MChA).

MChA corresponds to an entire class of effects in chiral media under an
external magnetic field, which show an enantio-selective difference in the
propagation of any unpolarized flux that propagates parallel or
anti-parallel to the magnetic field. This difference has its origin in the
simultaneous breaking of parity and time-reversal symmetries as a result of
the chirality of the media and the magnetization induced by the external
magnetic field, respectively. Generally, such a difference manifests itself
in the velocity or the attenuation of the flux. MChA has been predicted
since 1962 in the optical properties of chiral systems in magnetic fields 
\cite{groenewege61,burstein, baranova,wagniere,barron}, and was finally
observed in the 1990's \cite{Naturemca,kleindienst,mcaabs}. Nowadays it is
observed across the entire electromagnetic spectrum, from microwaves \cite%
{Microwave MChA} to X-rays \cite{Xray MChA}. The existence of MChA was
further generalized to electrical transport \cite{emchaprl} (in carbon nano
tubes \cite{eMChA CNT}, organic conductors \cite{Pop}, metals \cite%
{Yokouchi,Maurenbrecher,Aoki} and semiconductors \cite{Rikken Avarvari}), to
sound propagation \cite{Nomura} and to dielectric properties \cite{dMChA}.

EPR is basically a strongly resonant form of magnetic circular dichroism and
magnetic circular birefringence \cite{MCD review}, effects well known in the
optical wavelength range, where they however only represent small
perturbations of the optical properties of the medium. By analogy, one
should expect that MChA can manifest itself also in EPR of chiral media.
This expectation can be formalized by the observation that the EPR
transition probability $P$ induced by a propagating electromagnetic field
between the spin levels of a chiral medium in a magnetic field, is allowed
by parity and time-reversal symmetry to have the form%
\begin{equation}
P^{D/L}(\omega ,\hat{\mathbf{k}},\mathbf{B}_{0})=P_{0}(\omega
,B_{0})[1+\gamma ^{D/L}(\omega )\hat{\mathbf{k}}\cdot \mathbf{B}_{0}].
\label{Expansion}
\end{equation}%
In this equation, $\mathbf{B}_{0}$ is an external and constant magnetic
field, $P_{0}$ is the leading order transition probability between the
Zeeman levels, common to both enantiomers, the handedness of the medium is
represented by $D-$ right and $L-$ left, with $\gamma ^{D}=-\gamma ^{L}$,
and $\hat{\mathbf{k}}$ is a unitary vector in the direction of the wave
vector of the electromagnetic field driving the transition whose frequency $%
\omega $ is of the order of $\mu _{B}B_{0}/\hbar $. This shows that the EPR
transition probability is enantioselectively modified when probed by an
electromagnetic wave travelling parallel or anti-parallel to the magnetic
field, an effect that we shall call traveling wave enantioselective EPR
(TWEEPR). TWEEPR is quantified by the anisotropy factor $g_{T}^{D/L}$, which
represents the relative difference between the transition probabilities of
both enantiomers, 
\begin{equation}
g_{T}^{D/L}\equiv \frac{\lbrack P^{D/L}(\omega ,\widehat{\mathbf{k}},\mathbf{%
B}_{0})-P^{D/L}(\omega ,\widehat{\mathbf{k}},-\mathbf{B}_{0})]}{%
[P^{D/L}(\omega ,\widehat{\mathbf{k}},\mathbf{B}_{0})+P^{D/L}(\omega ,%
\widehat{\mathbf{k}},-\mathbf{B}_{0})]}=\gamma ^{D/L}\hat{\mathbf{k}}\cdot 
\mathbf{B}_{0}.  \label{gTEPR1}
\end{equation}%
As spin is related to the absence of time-reversal symmetry, and chirality
is related to the absence of parity symmetry, one might expect that the two
are decoupled and that $g_{T}^{D/L}$ is vanishingly small, thereby reducing
TWEEPR to an academic curiosity. However, below we will show through a model
calculation that, because of the ubiquitous spin-orbit coupling, TWEEPR
represents a significant and measurable fraction of the EPR transition
probability for realistic chiral systems and that its anisotropy factor is
not much smaller than that of optical MChA. Lastly, we will describe a
dedicated TWEEPR setup.

\noindent \emph{The model}\newline
As for the spin system of our model calculation of TWEEPR, without loss of
generality, we have chosen a crystalline quasi-octahedral Cu(II) chiral
complex because this ion is one of the most extensively studied systems by
EPR, it has the largest spin-orbit coupling among the first row transition
metals, and it has the simplest energy diagram. Its electromagnetic response
is attributed to a single unpaired electron that, in the $3d^{9}$
configuration of the Cu(II) complex, behaves as a hole of positive charge $%
+e $. We model the binding potential of the hole by that of an isotropic
harmonic oscillator that represents the rest of the ion, and is perturbed by
the chiral potential $V_{C}^{D/L}$ that results from its interaction with
the chiral environment of the crystal lattice, and by the spin-orbit
coupling. In turn, as we will show, this model allows us to find analytic
expressions for both the optical and the EPR magnetochiral anisotropy
parameters, $g_{O}^{D/L}$ and $g_{T}^{D/L}$, respectively, in terms of the
parameters of the model, both being proportional to the chiral coupling. Our
model can thus relate $g_{T}^{D/L}$ to its optical analogue $g_{O}^{D/L}$.
The latter is experimentally determined for several systems. In particular,
for CsCuCl$_{3}$ both MChD \cite{MChD CsCuCl3} and EPR \cite{EPR CsCuCl3}
have been reported. This approach thereby results in a generic analytical
expression for $g_{T}^{D/L}$ in terms of the parameters of our model, and in
a semi-empirical and quantitative prediction for $g_{T}^{D/L}$ for this
particular material in terms of its experimental optical MChD. The latter
can be extended to any material for which optical MChD has been determined.
Below we detail our model, which is a variant of Condon's model for optical
activity \cite{Condon,Condon2}, and its extension to optical magnetochiral
birefringence \cite{Donaire EJP}.\newline
The Hamiltonian describing the system is given by $%
H=H_{0}+V_{C}^{D/L}+V_{SO} $, with%
\begin{equation}
H_{0}=\frac{p^{2}}{2m_{e}}+\frac{m_{e}\omega _{0}^{2}r^{2}}{2}-\mu _{B}(%
\mathbf{L}+g\mathbf{S)}\cdot \mathbf{B}_{0},  \label{HO_hamiltonian}
\end{equation}%
\begin{equation}
V_{C}^{D/L}=C^{D/L}xyz,\quad V_{SO}=\lambda \mathbf{L}\cdot \mathbf{S},
\label{spin_orbit_hamiltonian}
\end{equation}%
where $\mathbf{r}=(x,y,z)$ and $\mathbf{p}$ are the position and kinetic
momentum vectors of the harmonic oscillator, $\omega _{0}$ is its natural
frequency, $\mathbf{L}$ and $\mathbf{S}$ are their
orbital and spin angular momentum operators, respectively, $C^{D}=-C^{L}$ is the right/left-handed chiral
coupling, $g\simeq 2$ is the Land\'{e} factor, $\lambda \simeq -0.1\,$eV is
the spin-orbit (SO) coupling parameter, and $\mathbf{B}_{0}\equiv B_{0}\hat{%
\mathbf{z}}$ is the external magnetic field. The interaction with an
electromagnetic plane-wave of frequency $\omega $, propagating along $%
\mathbf{B}_{0}$, is given in a multipole expansion by 
\begin{equation}
W=-e\mathbf{r}\cdot \mathbf{E}_{\omega }(t)/2-\mu _{B}(\mathbf{L}+g\mathbf{S)%
}\cdot \mathbf{B}_{\omega }(t)/2+\text{h.c.},  \label{Interaction}
\end{equation}%
where $\mathbf{E}_{\omega }(t)=i\omega \mathbf{A}_{\omega }e^{-i\omega t}$
and $\mathbf{B}_{\omega }(t)=i\bar{n}\mathbf{k}\wedge \mathbf{A}_{\omega
}e^{-i\omega t}$ are the complex-valued electric and magnetic fields in
terms of the electromagnetic vector potential, $\mathbf{A}_{\omega }$,
evaluated at the center of mass of the ion. Note that the field incident on
a molecule of the complex is the effective field which propagates throughout
the medium with an effective index of refraction $\bar{n}$. Hence it is the
effective wavevector $\bar{n}\mathbf{k}$ that appears. 
\begin{figure}[h]
\includegraphics[height=7.3cm,width=8.6cm,clip]{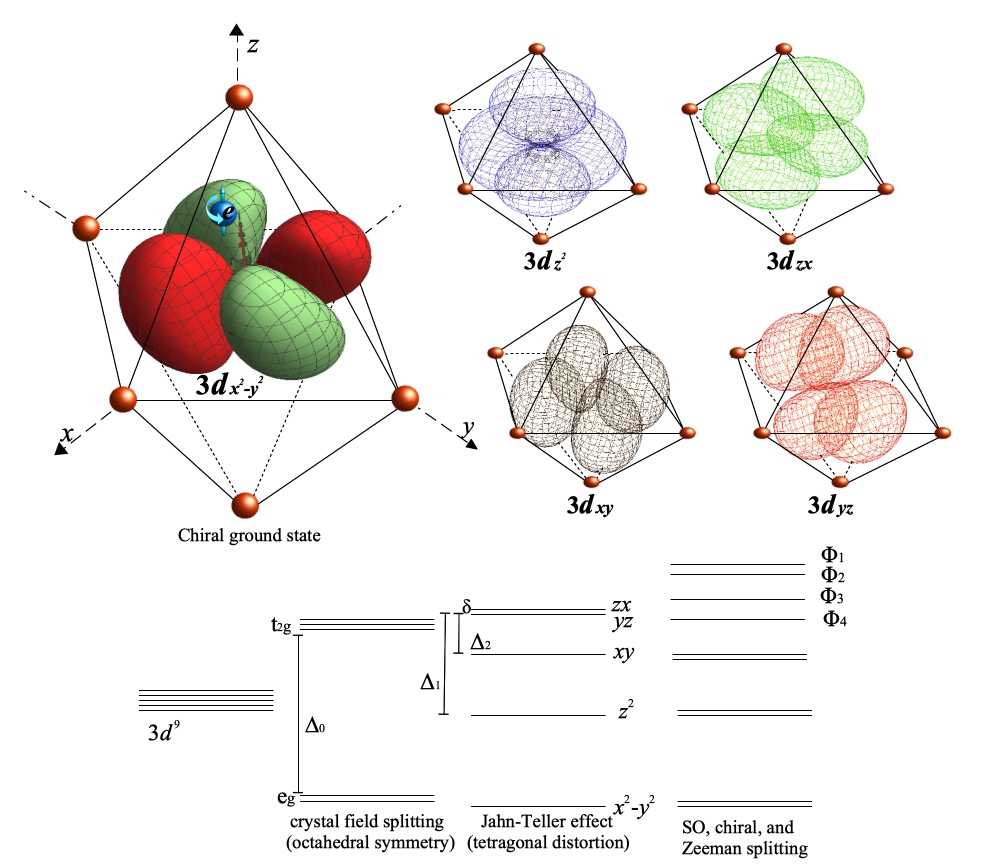}
\caption{Energy levels of Cu(II) in a chiral quasi-octahedral configuration.
Approximate experimental values are $\Delta _{0}\simeq 1.5$ eV, $\Delta
_{1}\simeq 0.5$ eV, $\Delta _{2}\simeq≈0.23$ eV.}
\label{Energy levels}
\end{figure}

In our model, the $3d$ orbitals are represented by linear combinations of
the $n=2$, $l=2$ states of the isotropic harmonic oscillator --see Appendix \ref{appA}.
Essential to the original Condon model was the anisotropy of the harmonic
oscillator, which removes all axis and planes of symmetry. In our model,
such an anisotropy is provided by the interaction of the ion with the
surrounding ligands of the complex, which in the case of CsCuCl$_{3}$ form
an quasi-octahedral structure. In the first place, that interaction causes
the elongation of the $3d$ orbitals which lie along the $z$-axis, opening an
optical gap $\Delta _{0}$. Also, in conjunction with the Jahn-Teller
distortion and the helical configuration of the Cu(II) ions, it removes the
degeneracy between the orbitals lying on the $xy$ plane and generates a
small energy gap $\delta $ between the states $d_{zx}$ and $d_{yz}$, with $%
\lambda \gg \delta $. The ground state of the Cu(II) ion in the octahedral
configuration $\Psi $ is, at finite temperature and subject to a magnetic
field, a linear combination of the doublet $d_{x^{2}-y^{2}}\otimes
\{\uparrow ,\downarrow \}$, 
\begin{equation}
|\Psi \rangle =|d_{x^{2}-y^{2}}\rangle \otimes (\cos {\theta /2}\uparrow
+\sin {\theta /2}\downarrow ),
\end{equation}%
where $\theta $, being a function of $B_{0}$ and the temperature, is the
angle between the magnetization of the sample and $\mathbf{B}_{0}$. For EPR,
spin-flip takes place at a resonance frequency $\Omega =g\mu _{B}B_{0}/\hbar 
$ when the up $\uparrow $ component of $\Psi $ turns into $|\Phi \rangle
=|d_{x^{2}-y^{2}}\rangle \otimes \downarrow $ , with probability
proportional to $\cos ^{2}\theta /2$, and the down $\downarrow $ component
turns into $|\Phi ^{\prime }\rangle =|d_{x^{2}-y^{2}}\rangle \otimes
\uparrow $ with probability proportional to $\sin ^{2}\theta /2$. The net
absorption probability is thus proportional to $\cos ^{2}\theta /2-\sin
^{2}\theta /2=\cos \theta $ and hence to the degree of magnetization along $%
\mathbf{B}_{0}$. At \ $B_{0}$ = 1T, $\Omega $ corresponds to an energy 150 $%
\mu $eV. In contrast, optical absorption happens at an energy $\Delta
_{0}\simeq 1.5$ eV towards the quadruplet $\{d_{zx},d_{yz}\}\otimes
\{\uparrow ,\downarrow \}$. Applying standard perturbation theory with the
spin-orbit and the Zeeman potentials upon this quasidegenerate quadruplet ,
we end up with the four states $\phi _{i}$, $i=1,..,4$, as appear in the
energy diagram represented in Fig.\ref{Energy levels} --a brief description
can be found in the Appendix \ref{appA}. It is of note that
these states play a crucial role in the E1M1 transitions of both EPR and its
optical analogue.

\noindent \emph{Results}\newline
Using up to fourth order time-dependent perturbation theory on $V_{SO}$, $%
V_{C}$ and $W$, in the adiabatic regime, our model allows us to
calculate the standard EPR and optical transition probabilities, as well as
the MChA corrections to both of them, with the latter two being both
proportional to $C^{D/L}$. As for $g_{T}^{D/L}$, the probability difference
in the denominator of Eq.(\ref{gTEPR1}) is an enantioselective E1M1
transition, whereas the denominator equals in good approximation the leading
order M1M1 transition, $g_{T}^{D/L}=P_{E1M1}^{D/L}/P_{M1M1}|_{\omega \approx
\Omega }$, with  
\begin{widetext}
\begin{align}
P_{M1M1}|_{\omega\approx\Omega}&=\hbar^{-2}\Bigl|\int_{0}^{\mathcal{T}}\text{d}t e^{-i(\mathcal{T}-t)(\Omega/2 -i\Gamma
/2)}e^{-it(\omega -\Omega/2 )}\langle \Phi |-g\mu _{B}\mathbf{S}\cdot \mathbf{B%
}_{\omega }|\Psi \rangle \Bigr|^{2}-\hbar^{-2}\Bigl|\int_{0}^{\mathcal{T}}\text{d}t e^{-i(\mathcal{T}-t)(2\omega -\Omega/2 -i\Gamma
/2)}\nonumber\\
&\times e^{-it(\omega +\Omega/2 )}\langle \Phi ^{\prime }|-g\mu _{B}\mathbf{S}%
\cdot \mathbf{B}_{\omega }|\Psi \rangle \Bigr|^{2},\nonumber\\
P_{E1M1}^{D/L}|_{\omega\approx\Omega}&=-2\hbar^{-2}\text{Re}\int_{0}^{\mathcal{T}}
\text{d}t e^{-i(\mathcal{T}-t)(\Omega/2 -i\Gamma /2)}\langle \tilde{\Phi} |-e\mathbf{r}\cdot(\bar{n}^{2}+2) 
\mathbf{E}_{\omega}/3|\tilde{\Psi}\rangle e^{-it(\omega
-\Omega/2 )}\int_{0}^{\mathcal{T}}\text{d}\tau\:e^{i(\mathcal{T}-\tau )(\Omega/2 +i\Gamma
/2)}  \nonumber \\
&\times\langle\Psi|-g\mu _{B}\mathbf{S}\cdot \mathbf{B}_{\omega }^{\ast
}|\Phi\rangle e^{i\tau (\omega -\Omega/2 )}+2\hbar^{-2}\text{Re}\int_{0}^{\mathcal{T}}\text{d}t\:e^{-i(\mathcal{T}-t)(2\omega -\Omega/2 )}\langle\tilde{\Phi}^{\prime}|-e\mathbf{r}\cdot 
(\bar{n}^{2}+2)\mathbf{E}_{\omega}/3|\tilde{\Psi}\rangle\nonumber\\
&\times e^{-it(\omega+\Omega/2 -i\Gamma /2)}\int_{0}^{\mathcal{T}}\text{d}\tau\:e^{i(\mathcal{T}-\tau )(2\omega-\Omega/2 )}
\langle\Psi|-g\mu _{B}\mathbf{S}\cdot \mathbf{B}_{\omega}^{\ast }|\Phi^{\prime}\rangle 
e^{i\tau (\omega +\Omega/2 +i\Gamma /2)},  \quad  \Gamma\mathcal{T}\gg 1,\label{perturbEPR}
\end{align}
\end{widetext}where $\Gamma $ is the linewidth of EPR absorption, $\Gamma 
\mathcal{T}\gg 1$ implies the adiabatic approximation, and the states $%
\tilde{\Psi}$, $\tilde{\Phi}$, and $\tilde{\Phi}^{\prime }$ are dressed with
the states $\phi _{i}$, $i=1,..,4$, on account of the spin-orbit and chiral
interactions. Using a linearly polarized microwave probe field in Eq.(\ref%
{perturbEPR}), the resultant expression for the TWEEPR anisotropy factor
reads 
\begin{equation}
g_{T}^{D/L}\simeq \frac{c\,C^{D/L}\hbar \,\Omega \,\delta }{m_{e}\omega
_{0}^{3}\Delta _{0}^{2}}\frac{\bar{n}^{2}+2}{3\bar{n}},  \label{g sub T}
\end{equation}%
where the second factor on the right hand side describes the effect of the
refractive index on the local electric field and the wavevector. It is worth
noting that the aforementioned dependence on magnetization, $\sim \cos {%
\theta }$, cancels out in the ratio between probabilities. For further
details, see Appendix \ref{appB}.

The values for the unknown parameters in Eq.(\ref{g sub T}) can be deduced
comparing the predictions of the model with the experimental results for
optical MChD \cite{MChD CsCuCl3} and EPR \cite{EPR CsCuCl3}\ in CsCuCl$_{3}$%
. In particular, we can estimate $g_{T}^{D/L}$ from the data on the
non-reciprocal absorption coefficient in optical MChD, $\alpha _{A}=\alpha (%
\mathbf{B}_{0}\upharpoonleft \upharpoonright \mathbf{k})-\alpha (\mathbf{B}%
_{0}\downharpoonleft \upharpoonright \mathbf{k})$. The calculation goes as
follows. In terms of the E1M1 absorption probability at resonance, $\omega
=\Delta _{0}/\hbar $, $\alpha _{A}$ reads 
\begin{equation}
\alpha _{A}=\frac{4c\mu _{0}\rho \Delta _{0}\Gamma ^{\prime }}{|E_{\omega
}|^{2}}P_{E1M1}^{D/L}|_{\omega =\Delta _{0}/\hbar },
\end{equation}%
where $\Gamma ^{\prime }$ is the linewidth of optical absorption, and $\rho $
is the molecular number density of the complex. Using our model, a
calculation analogous to that for $P_{E1M1}^{D/L,EPR}$ but for its optical
counterpart, $P_{E1M1}^{D/L,O}$ -- Appendices \ref{appB}, \ref{appC} and \ref{appD}-, allows as to
express $g_{T}^{D/L}$ in Eq.(\ref{g sub T}) in terms of $\alpha _{A}$, 
\begin{equation}
g_{T}^{D/L}=\frac{c\,\hbar ^{3}\Gamma ^{\prime }\Omega \tilde{\Delta}\alpha
_{A}}{2\Delta _{0}^{3}\mu _{0}\mu _{B}^{2}\rho \cos {\theta }},
\label{gTEPR}
\end{equation}%
where $\tilde{\Delta}^{-1}=\Delta _{0}^{-1}+\Delta _{2}^{-1}-3\Delta
_{1}^{-1}$ is the inverse of an effective energy interval which takes
account of the optical transitions to intermediate states --see Fig.\ref%
{Energy levels}. It is of note that, whereas the magnetic transition is
driven in EPR by the spin operator [Eq.(\ref{perturbEPR})], it is driven by
the orbital angular momentum in the optical case. In turn, this causes MChD
to be stronger in the optical case and proportional to the degree of
magnetization $\cos {\theta }$, which can be approximated by $\cos {\theta }%
\approx \mu _{0}B_{0}/k_{B}T$ \cite{magnetization}. The optical MChA
parameter, $g_{0}^{D/L}$, has an analogous expression to that in Eq.(\ref%
{gTEPR1}) with $\hbar \omega \approx \Delta _{0}$, being proportional to $%
\alpha _{A}$. Hence, our model allows us to estimate its upper bound, $%
g_{0}^{D/L}\leq (cC^{D/L}\delta \cos {\theta })/(m_{e}\omega _{0}^{3}\tilde{%
\Delta})$ -- see Appendices \ref{appC} and \ref{appD}, from which $g_{T}^{D/L}/g_{0}^{D/L}\gtrsim (\hbar
\Omega \tilde{\Delta})/(\Delta _{0}^{2}\cos {\theta })$. Note that, since
both $\Omega $ and $\cos {\theta }$ are proportional to $B_{0}$, the ratio
between EPR and optical MChA factors is independent of the field strength.

Finally, substituting the experimental values for CsCuCl$_{3}$ of all the
variables in Eq.(\ref{gTEPR}), for $B_{0}=14$ T at a temperature of 4.2 K,
we obtain $g_{T}^{D/L}\approx 1.5\cdot 10^{-2}$, which is small but not
beyond the resolution of high field EPR spectrometers. For an X band EPR
spectrometer ($B=0,35$ T), this means $g_{T}^{D/L}\approx 3\cdot 10^{-4\text{
}}$which will require a different approach, as we discuss below.

\medskip \noindent \emph{Implementation}\newline
In commercial EPR spectrometers, resonant standing wave cavities are used to
enhance sensitivity. Such a cavity can be regarded as containing equal
amounts of traveling waves with $\mathbf{k}$ and $-\mathbf{k.}$ The MChA $%
\gamma ^{D/L}$ term in Eq.(\ref{Expansion}) can therefore not give a net
contribution to the resonance in such a configuration. For this term to be
observed, a traveling wave configuration should be used. Such configurations
are not unknown in EPR; several reported home-built EPR spectrometers have
used one-pass transmission configurations \cite{Early Transmission EPR} \cite%
{Broadband EPR}. Sensitivity for such a travelling wave configuration can be
enhanced by means of a Mach-Zehnder interferometer \cite{Mach Zehnder} or a
unidirectional ring resonator \cite{Travelling resonator}. 
\begin{figure}[h]
\includegraphics[height=6.6cm,width=6.1cm,clip]{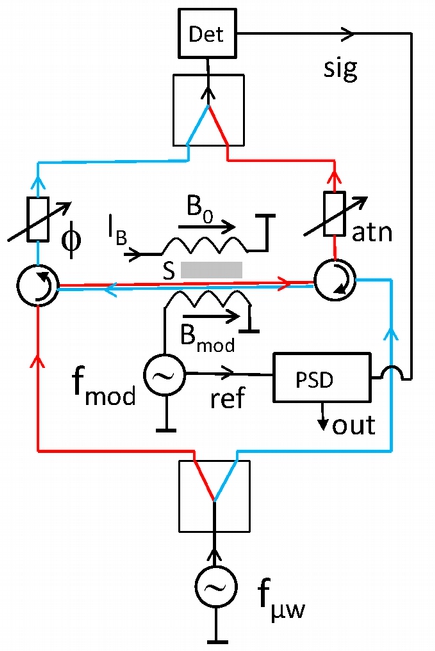}
\caption{Schematic setup of the TWEEPR interferometer. The waves
counterpropagating through the sample S are depicted in red and blue.}
\label{Figure 2}
\end{figure}
In such a configuration, MChA can be obtained as the difference between the
microwave transmissions for the two opposing magnetic field directions,
similar to what was realized in the optical case \cite{mcaabs}. As the EPR
lines can be quite narrow, the two oppositely oriented magnetic fields
should have the same magnitude with high precision, which requires a tight
control of this field, possibly with another EPR\ or NMR feedback circuit.
Stabilizing a field this way can be quite time-consuming, and TWEEPR being a
small difference on the already small EPR absorption, the extensive
signal-averaging through field alternations that would be required to obtain
a good signal-to-noise-ratio, makes such an approach impractical. We
therefore propose another approach in the form of an\ X band microwave
interferometer that removes the normal EPR contribution from the output
signal, through destructive interference between counter-propagating waves
through the sample at a fixed magnetic field, as illustrated in Figure \ref%
{Figure 2}. This leaves ideally only the
TWEEPR contribution. By applying an additional small modulation field and
using phase sensitive detection (PSD) sufficient sensitivity is obtained to
resolve this small contribution. When tuned to total destructive
interference at zero field, the interferometer output as given by the PSD is
proportional to the TWEEPR response $d[T(\mathbf{B}_{0}\upharpoonleft
\upharpoonright \mathbf{k})-T(\mathbf{B}_{0}\downharpoonleft \upharpoonright 
\mathbf{k})]/dB_{0}=\gamma ^{D/L}(\omega )$. The sensitivity of the
interferometer can be further improved by inserting the sample in a
unidirectional resonant ring resonator. Q factors above $10^{3}$ have been
reported for such configurations \cite{High Q ring} and would bring a
corresponding increase in sensitivity.\ It seems therefore quite feasible
that TWEEPR can evolve into a standard characterization technique in the
form of standalone dedicated TWEEPR spectrometers. An alternative to this
configuration could be the microwave equivalent of the first observation of
optical MChA in luminescence \cite{Naturemca}, using pulsed EPR echo
techniques \cite{EPR review} with a similar interferometer setup.

\medskip \noindent \emph{Discussion}\newline
In general, the non-local response of a chiral system of size $a$ to an
electromagnetic wave with wave vector $k$ is of the order $ka$, so one could
have expected $g_{T}^{D/L}/g_{O}^{D/L}$ to be of the order $\hbar \Omega
/\Delta _{0}$, the relevant spatial length scale for both TWEEPR and optical
MChD being the orbital size. This ratio is of the order of $10^{-4}$, which
would have put TWEEPR beyond experimental reach. However, in contrast to the
optical absorption, which to zeroth order is independent of the magnetic
field, the normal EPR absorption scales with the magnetization of the spin
system. Since the MChA corrections are proportional to the magnetization in
both EPR and the optical case, the cancellation of the factor $\cos {\theta
\ll 1}$ applies to $g_{T}^{D/L}$ only, and it appears thereby in the
denominator of $g_{T}^{D/L}/g_{O}^{D/L}$, resulting in Eq.(\ref{gTEPR}). For
room temperature X-band EPR of Cu(II), this results in $%
g_{T}^{D/L}/g_{O}^{D/L}$ of the order of $10^{-1}$, which makes TWEEPR
experimentally feasible under those conditions. As a consequence, and in
contrast to many other magnetic resonance techniques, going to low
temperatures is not necessarily favorable for TWEEPR. Going to higher
magnetic field does not affect $g_{T}^{D/L}/g_{O}^{D/L}$, the increase in $%
\Omega $ being compensated by the concomitant increase of $\cos {\theta }$
because of the higher resonance field.

The main results of our model are an analytic expression for the TWEEPR
anisotropy factor [Eq.(\ref{g sub T})] and an expression for its
relationship with the optical anisotropy absorption coefficient [Eq.(\ref%
{gTEPR})]. The expression in Eq.(\ref{g sub T}) shows that $g_{T}^{D/L}$ has
a linear dependence on the magnetic field strength (through $\Omega $) and
on the chirality (through $C^{D/L}$), as predicted by symmetry arguments.
The dependence on the spin-orbit coupling does not appear explicitly,
because we\ have considered the case for Cu(II), where the level splitting $%
\delta $ is much smaller than the SO coupling $\lambda $. In the inverse
case, $g_{T}^{D/L}$ would be proportional to $\lambda $ instead. Adapting
the calculation to other chiral transition metal complexes is conceptually
straightforward and should result in an expression similar to Eq.(\ref{g sub
T}), apart from numerical factors of order unity. A rather different case is
represented by chiral organic radicals, where the unpaired electron is
delocalized on one or more interatomic bonds and a different microscopic
model should be used for the calculation of $g_{T}^{D/L}$. One might however
expect that such differences apply also to the calculation of $g_{O}^{D/L}$
for such radicals, preserving a relationship similar to that in Eq.(\ref%
{gTEPR}).

\noindent \emph{Acknowledgements}\newline
This work was supported by the Agence Nationale de la Recherche (SECRETS,
(ANR PRC 20-CE06-0023-01) and the Laboratory of Excellence NanoX
(ANR-17-EURE-0009)). We gratefully acknowledge helpful discussions with
Anne-Laure Barra.

\appendix

\begin{widetext}
In the Appendices we describe the theoretical model used in our calculations, 
we offer explicit expressions for the transition probabilities that enter the anisotropy factors in EPR and 
optical MChD, and comment on the limitations of our model.

\section{Fundamentals of the model}\label{appA}
As outlined in the article, in order to estimate the MChA factors of a chiral Cu(II) complex, we consider a 
variant of the one-electron model proposed by Condon 
for the study of natural optical activity in chiral compounds \cite{Condon,Condon2}.  
The total Hamiltonian of our model is $H=H_{0}+V_{C}^{D/L}+V_{SO}$, where $H_{0}=\frac{p^{2}}{2m_{e}}+\frac{m_{e}\omega _{0}^{2}r^{2}}{2}+V_{Z}$ is the unperturbed Hamiltonian, 
with $V_{Z}=-\mu _{B}(\mathbf{L}+g\mathbf{S)}\cdot \mathbf{B}_{0}$ being the Zeeman potential; and 
$V_{C}^{D/L}=C^{D/L}xyz,\: V_{SO}=\lambda \mathbf{L}\cdot \mathbf{S}$ being the chiral potential 
and the spin-orbit coupling, respectively.  We stick to the nomenclature used in the article. 
 The chiral Hamiltonian, $V_{C}^{D/L}$, results from the electrostatic  
interaction of the ion with the chiral configuration of the ligands in the complex, and produces
the necessary parity asymmetry which is at the origin of natural optical activity.  
The orbital contribution of the Zeeman potential was added in Ref.\cite{Donaire EJP} to the original Condon's model to estimate the magneto-chiral birefringence of
diamagnetic chiral compounds. In order to account for magnetochiral dichroism (MChD) in a paramagnetic complex, we introduce here the spin contribution to the Zeeman
potential as well as the spin-orbit coupling. In contrast to
the approach in Ref.\cite{Donaire EJP} and for simplicity, we consider an isotropic harmonic oscillator, whereas 
the anisotropy caused by the crystal field is introduced in an effective manner through the energy intervals between the $3d$ orbitals, 
as depicted in Fig.1 in the article.

The eigenstates of $H_{0}$ are labeled with the eigenvalues of the orbital angular momentum and spin operators, $%
\{|n_{L},n_{R},n_{z}\rangle \}\otimes \{\uparrow ,\downarrow \}$ \cite%
{Cohenbookvol2}, upon which $V_{C}^{D/L}$ and $V_{SO}$ act perturbatively. In a Cu(II) complex, the chromophoric charge is the unpaired electron of the 
$3d^{9}$ electronic configuration which behaves as a hole of positive charge. In the absence of ligands, the $3d$
orbitals of the ion can be represented approximately by the $n=2$, $l=2$
states of the harmonic oscillator of our model. However, the ligands'
fields affect the electronic configuration of the ion, removing the
degeneracy of the $d$-states. In particular, for octahedral coordination
geometries around the ion, the set of $d$-orbitals splits into doubly
degenerate e$_{g}$ orbitals, $d_{x^{2}-y^{2}}$ and $d_{z^{2}}$, and triply
degenerate t$_{2g}$ orbitals, $d_{xy}$, $d_{yz}$ and $d_{zx}$. The energy
interval between e$_{g}$ and t$_{2g}$ states, $\Delta _{0}$, lies in the
visible region of the spectrum, $\Delta _{0}\simeq 1.5$ eV. As a result, the e%
$_{g}$ orbitals become the ground states, and can be approximated by
linear combinations of $l=2$, $m_{l}=0,\pm 2$ eigenstates of the harmonic
oscillator. The fact that the chromophoric charge in the e$_{g}$ states
cannot rotate into any other orbital leads to an effective quenching of the
orbital angular momentum of the ground state. Below a certain temperature, an additional Jahn-Teller (JT) distortion takes
place when the ligands along one of the axes, say the $z$-axis, move away
from the ion in order to minimize the electronic repulsion, giving rise to
the complete removal of the degeneracy in the e$_{g}$ level, and to a
partial lifting of the degeneracy in the t$_{2g}$ orbitals. The
isotropy of the system is thus broken and the ground state becomes unique, up
to spin degeneracy. For the particular case of the CsCuCl$_{3}$ crystal, the
bonds along the $z$-axis get elongated and the ground state is the $%
d_{x^{2}-y^{2}}$ orbital. Fig.1 in the article depicts
the energy splitting of the distorted $d$-orbitals, including the
approximate values of the energy intervals. Lastly, the JT distortion in 
conjuntion with the helical deformation of the crystal along the $c$-axis, of coordiates [1,1,1] in the 
local axis basis, removes the degeneracy
between the orbitals lying on the $xy$ plane in a small ammount $\delta$. 
Below, we write the approximate expression of the $3d$ orbitals in terms of 
the harmonic oscillator eigenstates, $\{|n_{L},n_{R},n_{z}\rangle \}$, together with their corresponding energies,
\begin{align}
&|d_{zx}\rangle=(|0,1,1\rangle-|1,0,1\rangle)/\sqrt{2},\quad\mathcal{E}=\Delta_{0},\nonumber\\
&|d_{yz}\rangle=i(|0,1,1\rangle+|1,0,1\rangle)/\sqrt{2},\quad\mathcal{E}=\Delta_{0}-\delta,\nonumber\\
&|d_{xy}\rangle=i(|0,2,0\rangle-|2,0,0\rangle)/\sqrt{2},\quad\mathcal{E}=\Delta_{0}-\Delta_{2},\nonumber\\
&|d_{z^{2}}\rangle=(|1,1,0\rangle-\sqrt{2}|0,0,2\rangle)/\sqrt{3},\quad\mathcal{E}=\Delta_{0}-\Delta_{1},\nonumber\\
&|d_{x^{2}-y^{2}}\rangle=(|0,2,0\rangle+|2,0,0\rangle)/\sqrt{2},\quad\mathcal{E}=0.
\end{align}
Altogether, the crystal field combined with the JT distortion 
and the helical deformation turns the crystalline structure into a chiral
one. In accord with Condon's model, the potential $V^{D/L}_{C}$ reproduces the
electrostatic interaction of the chromophoric charge with the surrounding
chiral structure, removing all axes and planes of symmetry from the system.
It is through the chiral potential that E1 transitions between the $3d$ orbitals take place in
our model. In addition to the above interactions, MChD in EPR requires
necessarily the coupling between the spin and the orbital angular momentum
of the unpaired electron hole through the potential $V_{SO}$, where the coupling constant is $\lambda \approx -0.1$
eV. In particular, the SO interaction together with the Zeeman potential break the quasi-degeneracy between the four 
states $\{|d_{zx}\rangle,|d_{yz}\rangle\}\otimes \{\uparrow ,\downarrow \}$, providing the following eigenstates for 
$\lambda\gg\delta$, 
\begin{align}
|\Phi _{1}\rangle & \approx |1,0,1\rangle \otimes \downarrow +\frac{\delta}{2\lambda} |0,1,1\rangle \otimes \downarrow ,\quad  \notag \nonumber\\
\mathcal{E}& \simeq \Delta _{0}-\lambda /2+\hbar \Omega, \nonumber\\
|\Phi _{2}\rangle & \approx |0,1,1\rangle \otimes \uparrow +\frac{\delta}{2\lambda}
|1,0,1\rangle \otimes \uparrow ,\quad  \notag \nonumber\\
\mathcal{E}& \simeq \Delta _{0}-\lambda /2-\hbar \Omega,\nonumber\\
|\Phi _{3}\rangle & \approx |0,1,1\rangle \otimes \downarrow -\frac{\delta}{2\lambda}|1,0,1\rangle \otimes \downarrow ,\quad  \notag \nonumber\\
\mathcal{E}& \simeq \Delta _{0}+\lambda /2+\hbar \Omega+\frac{\delta ^{2}}{4\lambda
^{2}}(\lambda +\hbar \Omega ), \nonumber\\
|\Phi _{4}\rangle & \approx |1,0,1\rangle \otimes \uparrow -\frac{\delta}{2\lambda}
|0,1,1\rangle \otimes \uparrow ,\quad  \notag \nonumber\\
\mathcal{E}& \simeq \Delta _{0}+\lambda /2-\hbar \Omega+\frac{\delta ^{2}}{4\lambda
^{2}}(\lambda -\hbar \Omega ).\label{4phi}
\end{align}
$\{\Phi_{1},\Phi_{2},\Phi_{3},\Phi_{4}\}$ are indeed the eigenstates of the Hamiltonian $V_{Z}+V_{SO}$ restricted to the subspace 
$\{|d_{zx}\rangle,|d_{yz}\rangle\}\otimes \{\uparrow ,\downarrow \}$. They constitute the intermediate states of the transition processes in EPR 
mediated by the interaction of the spin with the chiral structure of the surrounding charges.

In the following, we apply to our system time-dependent quantum perturbation techniques to compute
first the MChA factor in EPR, $g_{T}^{D/L}$. Next, in order to estimate the
value of the unknowns of our model, we compute the anisotropy
factor in optical MChD for the same system. Finally, making use of the experimental values 
available for CsCuCl$_{3}$ in the literature \cite{EPR CsCuCl3,MChD CsCuCl3}, we estimate 
the strength of TWEEPR.

\section{MChD in EPR}\label{appB}
Let us consider a CsCuCl$_{3}$ complex, initially prepared in its ground
state, and partially polarized along a uniform magnetic field $\mathbf{B}=B_{0}\hat{\mathbf{z}}$ directed along the $z$-axis, 
\begin{equation}
|\Psi \rangle =|d_{x^{2}-y^{2}}\rangle \otimes (\cos {\theta /2}\uparrow
+\sin {\theta /2}\downarrow )\approx \frac{1}{\sqrt{2}}(|0,2,0\rangle
+|2,0,0\rangle )\otimes (\cos {\theta /2}\uparrow +\sin {\theta /2}%
\downarrow ),
\end{equation}
where we have approximated the actual ground state with the 
corresponding state of our harmonic oscillator model in the basis $%
\{|n_{L},n_{R},n_{z}\rangle \}\otimes \{\uparrow ,\downarrow \}$, and $%
\theta $ is the angle between the magnetic moment of the complex and the $z$%
-axis, $\cos {\theta }=\hbar^{-1}\langle \Psi |2\mathbf{S}|\Psi \rangle \cdot \hat{%
\mathbf{z}}$. At temperature $T$, $\cos{\theta }\approx \mu _{0}B_{0}/k_{B}T$ \cite{magnetization}. Under the action of an incident electromagnetic field of
frequency $\omega $ close to the transition frequency, $\Omega =g\mu
_{B}B_{0}/\hbar $, and wave vector $\mathbf{k}$ parallel to $\mathbf{B}_{0}$%
, the complex gets partially excited towards the state 
\begin{equation}
|\Phi \rangle =|d_{x^{2}-y^{2}}\rangle \otimes \downarrow \approx \frac{1}{%
\sqrt{2}}(|0,2,0\rangle +|2,0,0\rangle )\otimes \downarrow ,
\end{equation}%
with probability proportional to $\cos ^{2}{\theta /2}$; and partially
de-excited (through stimulated emission) towards the state 
\begin{equation}
|\Phi ^{\prime }\rangle =|d_{x^{2}-y^{2}}\rangle \otimes \uparrow \approx 
\frac{1}{\sqrt{2}}(|0,2,0\rangle +|2,0,0\rangle )\otimes \uparrow ,
\end{equation}%
with probability proportional to $\sin ^{2}{\theta /2}$. Since the rest of 
probability factors are equivalent, the net absorption probability in
EPR is proportional to $\cos ^{2}{\theta /2}-\sin ^{2}{\theta /2}=\cos {%
\theta }$, and thus proportional to the magnetization of the
complex.

As mentioned in the article, from symmetry considerations and in leading order, the numerator and the
denominator in the ratio $g_{T}^{D/L}=[P^{D/L}(\omega ,\hat{\mathbf{k}},\mathbf{B}%
_{0})-P^{D/L}(\omega ,\hat{\mathbf{k}},-\mathbf{B}_{0})]/[P^{D/L}(\omega ,\hat{\mathbf{k}}
,\mathbf{B}_{0})+P^{D/L}(\omega ,\hat{\mathbf{k}},-\mathbf{B}_{0})]$ for $\omega\approx\Omega$ are
dominated, respectively, by the electric-magnetic dipole (E1M1) and the
magnetic-magnetic dipole (M1M1) transition probabilities, the magnetic
transition being driven by the spin operator only. That leads to the
approximate expression, 
\begin{equation}
g_{T}^{D/L}\simeq\frac{P_{E1M1}^{D/L}(\omega ,\hat{\mathbf{k}},\mathbf{B}_{0})}{P_{M1M1}(\omega ,\hat{\mathbf{k}},\mathbf{B}_{0})}\Bigr|_{\omega\approx\Omega}.
\end{equation}

In what follows, we compute the transition probabilities $P_{M1M1}$ and $P_{E1M1}^{D/L}$ for $\omega\approx\Omega$
using time-dependent perturbation theory in the adiabatic regime. This regime is the suitable one for a probe field whose duration is much longer than 
the typical lifetime for excitation or de-excitation. As in the article, the Hamiltonian of the interaction 
of our system with the microwave probe field reads, in the electric and magnetic dipole approximation, $W=-e\mathbf{r}\cdot \mathbf{E}_{\omega
}(t)/2-\mu _{B}(\mathbf{L}+2\mathbf{S})\cdot \mathbf{B}_{\omega }(t)/2+$%
h.c.. In this equation, $\mathbf{E}%
_{\omega }(t)=\mathbf{E}_{\omega }e^{-i\omega t}=i\omega \mathbf{A}_{\omega
}e^{-i\omega t}$, $\mathbf{B}_{\omega }(t)=\mathbf{B}_{\omega }e^{-i\omega
t}=i\bar{n}\mathbf{k}\wedge \mathbf{A}_{\omega }e^{-i\omega t}$, are the complex-valued electric and magnetic 
fields, respectively, with $%
\mathbf{A}_{\omega }$ being the complex-valued amplitude of the plane-wave
electromagnetic vector potential of frequency $\omega\approx\Omega$, evaluated at the center of mass of the
Cu(II) ion, and $\bar{n}$ being the effective refractive index of the sample. The local depolarization
changes the local electric field incident on each Cu(II) ion to $\mathbf{E}_{\omega }(\bar{n}^{2}+2)/3$.
Under the action of $W$, with $\mathbf{k}$ along $\mathbf{B}_{0}$, the
expressions for $P_{M1M1}$ and $P_{E1M1}^{D/L}$ read, respectively,
at leading order in the coupling constants of the interaction potentials,%
\begin{align}
P_{M1M1}|_{\omega\approx\Omega}& =\hbar^{-2}\left\vert \int_{0}^{\mathcal{T}}\text{d}t e^{-i(\mathcal{T}-t)(\Omega/2 -i\Gamma
/2)}e^{-it(\omega -\Omega/2 )}\langle \Phi |-g\mu _{B}\mathbf{S}\cdot \mathbf{B%
}_{\omega }|\Psi \rangle \right\vert ^{2}  \notag \\
& -\hbar^{-2}\left\vert \int_{0}^{\mathcal{T}}\text{d}t e^{-i(\mathcal{T}-t)(2\omega -\Omega/2 -i\Gamma
/2)}e^{-it(\omega +\Omega/2 )}\langle \Phi ^{\prime }|-g\mu _{B}\mathbf{S}%
\cdot \mathbf{B}_{\omega }|\Psi \rangle \right\vert ^{2},
\end{align}%
\begin{align}
P_{E1M1}^{D/L}|_{\omega\approx\Omega}& =2\text{Re}(-i)^{3}\hbar^{-4}\sum_{p,q\neq \Psi }\int_{0}^{\mathcal{T}}%
\text{d}t e^{-i(\mathcal{T}-t)(\Omega/2 -i\Gamma /2)}\langle \Phi |-e\mathbf{r}\cdot(\bar{n}^{2}+2) 
\mathbf{E}_{\omega}/3|p\rangle \int_{-\infty }^{t}\text{d}t^{\prime} e^{\eta
t^{\prime }}e^{-i(t-t^{\prime })(\mathcal{E}_{p}+\omega )}   \notag \\
& \times\langle
p|V^{D/L}_{C}|q\rangle \int_{-\infty }^{t^{\prime }}\text{d}t^{\prime \prime}e^{ \eta
t^{\prime \prime }}e^{-i(t^{\prime }-t^{\prime \prime })(\mathcal{E}%
_{q}+\omega )}\langle q|V_{SO}|\Psi \rangle e^{-it^{\prime \prime }(\omega
-\Omega/2 )}i\int_{0}^{\mathcal{T}}\text{d}\tau\:e^{i(\mathcal{T}-\tau )(\Omega/2 +i\Gamma
/2)}  \notag \\
&\times\langle\Psi |-g\mu _{B}\mathbf{S}\cdot \mathbf{B}_{\omega }^{\ast
}|\Phi \rangle e^{i\tau (\omega -\Omega/2 )}\:+2\text{Re}(-i)^{3}\hbar^{-4}\sum_{p,q\neq \Phi }\int_{-\infty }^{\mathcal{T}}\text{d}%
t\:e^{\eta t}e^{-i(\mathcal{T}-t)(\Omega/2 -i\Gamma /2)}\langle \Phi |V_{SO}|p\rangle  \notag \\
& \times\int_{-\infty }^{t}\text{d}t^{\prime}e^{\eta t^{\prime }}e^{-i(t-t^{\prime })%
\mathcal{E}_{p}}\langle p|V^{D/L}_{C}|q\rangle \int_{0}^{t^{\prime }}\text{d}t^{\prime \prime}e^{-i(t^{\prime
}-t^{\prime \prime })\mathcal{E}_{q}}\langle q|-e\mathbf{r}\cdot(\bar{n}^{2}+2) 
\mathbf{E}_{\omega}/3|\Psi \rangle e^{-it^{\prime \prime }(\omega -\Omega/2
)}\notag \\
&\times i\int_{0}^{\mathcal{T}}\text{d}\tau \:e^{i(\mathcal{T}-\tau )(\Omega/2 +i\Gamma /2)}\langle
\Psi |-g\mu _{B}\mathbf{S}\cdot \mathbf{B}_{\omega }^{\ast }|\Phi \rangle
e^{i\tau (\omega -\Omega/2 )}\notag \\
&-2\text{Re}(-i)^{3}\hbar^{-4}\sum_{p,q\neq \Psi }\int_{0}^{\mathcal{T}}\text{d}%
t\:e^{-i(\mathcal{T}-t)(2\omega -\Omega/2 )}\langle \Phi ^{\prime }|-e\mathbf{r}\cdot 
(\bar{n}^{2}+2)\mathbf{E}_{\omega}/3|p\rangle \int_{-\infty }^{t}\text{d}t^{\prime}e^{\eta
t^{\prime }}e^{-i(t-t^{\prime })(\mathcal{E}_{p}+\omega )}\notag \\
& \times \langle
p|V^{D/L}_{C}|q\rangle  \int_{-\infty }^{t^{\prime }}\text{d}t^{\prime \prime}e^{\eta
t^{\prime \prime }}e^{-i(t^{\prime }-t^{\prime \prime })(\mathcal{E}%
_{q}+\omega )}\langle q|V_{SO}|\Psi \rangle e^{-it^{\prime \prime }(\omega
+\Omega/2 -i\Gamma /2)}i\int_{0}^{\mathcal{T}}\text{d}\tau\:e^{i(\mathcal{T}-\tau )(2\omega
-\Omega/2 )}\notag \\
&\times\langle \Psi |-g\mu _{B}\mathbf{S}\cdot \mathbf{B}_{\omega
}^{\ast }|\Phi^{\prime}\rangle e^{i\tau (\omega +\Omega/2 +i\Gamma /2)}\:-2\text{Re}(-i)^{3}\hbar^{-4}\sum_{p,q\neq \Phi ^{\prime }}\int_{-\infty }^{\mathcal{T}}\text{d%
}t\:e^{\eta t}e^{-i(\mathcal{T}-t)(2\omega -\Omega/2 )}\notag \\
& \times\langle \Phi ^{\prime
}|V_{SO}|p\rangle \int_{-\infty }^{t}\text{d}t^{\prime}e^{\eta t^{\prime
}}e^{-i(t-t^{\prime })(2\omega +\mathcal{E}_{p})}\langle p|V^{D/L}_{C}|q\rangle \int_{0}^{t^{\prime }}\text{d}t^{\prime \prime } 
 e^{-i(t^{\prime }-t^{\prime \prime })(2\omega +\mathcal{E}%
_{q})} \notag \\
&\times\langle q|-e\mathbf{r}\cdot(\bar{n}^{2}+2) 
\mathbf{E}_{\omega}/3|\Psi \rangle
e^{-it^{\prime \prime }(\omega +\Omega/2 -i\Gamma /2)}i\int_{0}^{\mathcal{T}}\text{d}%
\tau\:e^{i(\mathcal{T}-\tau )(2\omega -\Omega/2 )}\langle\Psi |-g\mu _{B}\mathbf{S}%
\cdot \mathbf{B}_{\omega }^{\ast }|\Phi^{\prime}\rangle \notag \\
&\times e^{i\tau (\omega +\Omega/2
+i\Gamma /2)},  \quad \eta \rightarrow 0^{+},\:\: \Gamma \mathcal{T}\gg 1.
\end{align}%
In these equations the states $p$ and $q$ stand for the excited states of the $3d^{9}$ configuration together with other eigenstates of $H_{0}$ with $n\neq2$.
The quasi-stationary condition  $\eta\rightarrow0^{+}$ accounts for the stationarity of the chiral and the spin-orbit interactions; whereas the adiabatic 
limit $\Gamma\mathcal{T}\gg 1$ takes into account the long duration of the probe field with respect to the lifetime $\Gamma^{-1}$, with $\Gamma$ being the
linewidth of absorption and $\mathcal{T}$ the observation time. The diagrammatical representation of the processes involved in the above
equation is given in Fig.\ref{DiagramsEPR}. 
\begin{figure}[h]
\includegraphics[height=12.3cm,width=13.3cm,clip]{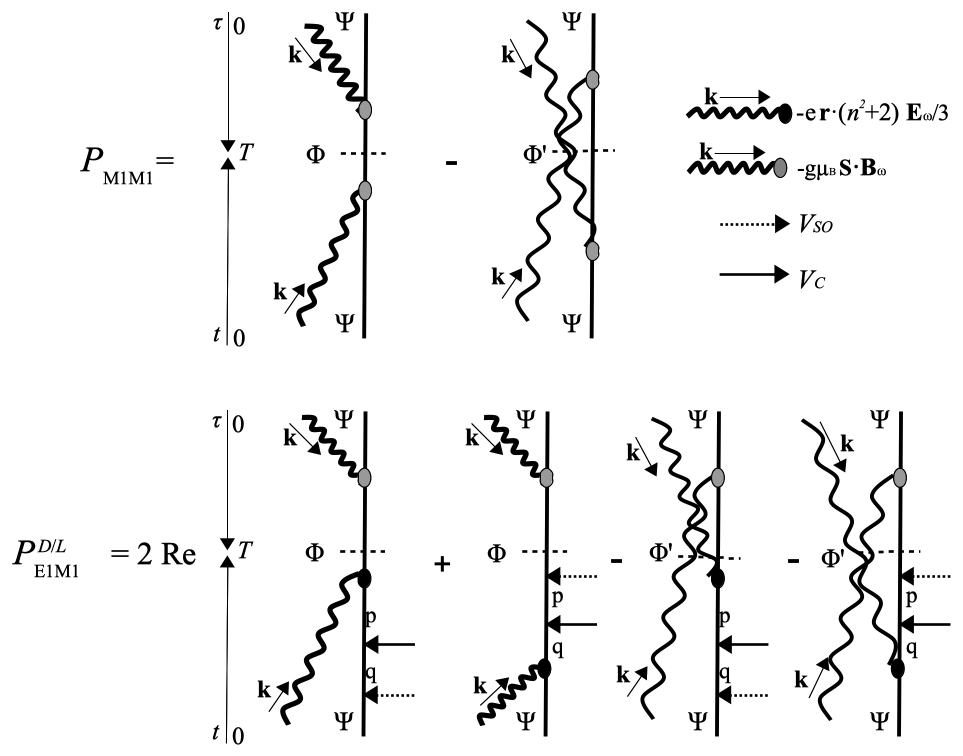}
\caption{Diagrammatic
representation of the processes which contribute to $P_{M1M1}$ and $%
P_{E1M1}^{D/L}$ for $\omega\approx\Omega$ at leading order in the perturbative interactions, i.e.,
at second order and fourth order, respectively. Time runs along the vertical
direction from 0 to the observation time $\mathcal{T}$, where the probability is
computed. Intermediate atomic states are labeled as $p$ and $q$. Diagrams with two-photon states account for stimulated emission.}\label{DiagramsEPR}
\end{figure}
In the article, the contributions of the quasi-stationary processes were incorporated into the 
dressed states $\tilde{\Psi}$, $\tilde{\Phi}$, $\tilde{\Phi}'$. More specifically, the bare states are dressed with the quadruplet $\{\Phi_{1},..,\Phi_{4}\}$ 
through $V_{SO}$, and with harmonic states with $n\neq2$ by $V_{C}$. In terms of the eigenstates of the harmonic oscillator, they read
\begin{align}
|\tilde{\Phi}\rangle&=\Bigr[(|020\rangle+|200\rangle)/\sqrt{2}+\frac{\lambda}{\sqrt{2}\Delta_{0}}(1+\Delta_{2}/\Delta_{0})(|020\rangle-|200\rangle)+\frac{i\lambda C^{D/L}K^{3/2}}{2\hbar\omega_{0}\Delta_{0}}(1+\Delta_{2}/\Delta_{0})\nonumber\\
&\times(|001\rangle-2|111\rangle)\Bigr]\downarrow\:+\:\Bigr[\frac{\lambda}{\sqrt{2}\Delta_{0}}(1+3\hbar\Omega/\Delta_{0})|011\rangle+\frac{\delta}{\sqrt{2}\Delta^{2}_{0}}(\hbar\Omega+\lambda/2)|101\rangle\nonumber\\
&+\frac{-iC^{D/L}K^{3/2}\lambda}{2\hbar\omega_{0}\Delta_{0}}(1+3\hbar\Omega/\Delta_{0})(|210\rangle-\sqrt{3}|030\rangle-\sqrt{2}|100\rangle)+\frac{iC^{D/L}K^{3/2}\delta}{2\hbar\omega_{0}\Delta^{2}_{0}}\nonumber\\
&\times(\hbar\Omega+\lambda/2)(|120\rangle-\sqrt{3}|300\rangle-\sqrt{2}|010\rangle)\Bigr]\uparrow\nonumber\\
|\tilde{\Phi}'\rangle&=\Bigr[(|020\rangle+|200\rangle)/\sqrt{2}+\frac{-\lambda}{\sqrt{2}\Delta_{0}}(1+\Delta_{2}/\Delta_{0})(|020\rangle-|200\rangle)+\frac{-i\lambda C^{D/L}K^{3/2}}{2\hbar\omega_{0}\Delta_{0}}(1+\Delta_{2}/\Delta_{0})\nonumber\\
&\times(|001\rangle-2|111\rangle)\Bigr]\uparrow\:+\:\Bigr[\frac{-\lambda}{\sqrt{2}\Delta_{0}}(1-3\hbar\Omega/\Delta_{0})|101\rangle+\frac{\delta}{\sqrt{2}\Delta^{2}_{0}}(\hbar\Omega-\lambda/2)|011\rangle\nonumber\\
&+\frac{-iC^{D/L}K^{3/2}\lambda}{2\hbar\omega_{0}\Delta_{0}}(1-3\hbar\Omega/\Delta_{0})(|120\rangle-\sqrt{3}|300\rangle-\sqrt{2}|010\rangle)+\frac{-iC^{D/L}K^{3/2}\delta}{2\hbar\omega_{0}\Delta^{2}_{0}}\nonumber\\
&\times(\hbar\Omega-\lambda/2)(|210\rangle-\sqrt{3}|030\rangle-\sqrt{2}|100\rangle)\Bigr]\downarrow\nonumber\\
|\tilde{\Psi}\rangle&=\cos{\theta/2}|\tilde{\Phi}'\rangle + \sin{\theta/2}|\tilde{\Phi}\rangle,\qquad K=\hbar/(2m_{e}\omega_{0}).
\end{align}

Using a linearly polarized incident field and averaging 
in orientations around the $\hat{\mathbf{z}}$-axis, we obtain, for $%
\lambda\gg\delta$, 
\begin{align}
P_{M1M1}|_{\omega\approx\Omega}& \simeq \frac{\hbar ^{-2}\mu _{B}^{2}|B_{\omega }|^{2}}{%
4[(\omega -\Omega )^{2}+\Gamma ^{2}/4]}\cos {\theta },  \label{Pm1m1EPR} \\
P_{E1M1}^{D/L}|_{\omega\approx\Omega}& \simeq \frac{(\bar{n}^{2}+2)}{3}\frac{C^{D/L}\Omega \delta }{m_{e}\omega
_{0}^{3}\Delta _{0}^{2}}\frac{\hbar ^{-1}\mu _{B}^{2}|B_{\omega }||E_{\omega
}|}{4[(\omega -\Omega )^{2}+\Gamma ^{2}/4]}\cos {\theta },  \label{Pe1m1EPR}
\\
g_{T}^{D/L}& \simeq \frac{(\bar{n}^{2}+2)}{3\bar{n}}\frac{c\,C^{D/L}\hbar \Omega \delta }{m_{e}\omega
_{0}^{3}\Delta _{0}^{2}}+\mathcal{O}(\delta
/\lambda ,\lambda /\Delta _{0}).  \label{AgTEPR}
\end{align}%
Lastly, it is worth mentioning that for the case $\delta >\lambda $, i.e.,
when anisotropy dominates over the spin-orbit coupling, $g_{T}^{D/L}$ scales
as $(c\hbar C^{D/L}\Omega \delta \lambda) /(m_{e}\omega _{0}^{3}\Delta _{0}^{3})$
instead. This scenario will be addressed in a separate publication \cite{inpreparation}.

\section{Optical MChD}\label{appC}
Optical MChD  involves transitions of frequency $\Delta _{0}$
from the ground state $|\Psi \rangle $ to the quasi-degenerate quadruplet $\{|d_{zx}\rangle,|d_{yz}\rangle\}\otimes \{\uparrow ,\downarrow \}$ which, in account of 
the Zeeman and spin-orbit interactions, for $\delta \ll\lambda$, corresponds to the set of states $\{\Phi_{1},...,\Phi_{4}\}$ of Eq.(\ref{4phi}).  
In contrast to EPR, the absorption probability in the denominator of the
ratio $g_{O}^{D/L}=[P^{D/L}(\omega ,\hat{\mathbf{k}},\mathbf{B}%
_{0})-P^{D/L}(\omega ,\hat{\mathbf{k}},-\mathbf{B}_{0})]/[P^{D/L}(\omega ,\hat{\mathbf{k}}
,\mathbf{B}_{0})+P^{D/L}(\omega ,\hat{\mathbf{k}},-\mathbf{B}_{0})]$ for $\omega\approx\Delta_{0}/\hbar$ may not be dominated by
the magnetic-magnetic dipole absorption probability. This might be
so because the $d$-orbitals of the Cu(II) ion hybridize generally with the $%
\sigma $ and $\pi $ orbitals of the ligands, allowing for additional
electric-electric dipole (E1E1) transitions. For the sake of
simplicity, we will neglect the latter in our calculations, which implies
that our preliminar estimate for $g_{O}^{D/L}$  must be intended as an
approximate upper bound. 
As for the case of EPR, the numerator of the ratio in $%
g_{O}^{D/L}$ is again dominated by the electric-magnetic dipole absorption
probability, and the non-vanishing terms come from magnetic transitions driven by the spin angular momentum --Eq.(\ref{Pe1m1Opt}) below. 
However, in contrast to EPR, the magnetic transitions in the denominator are mainly driven by the orbital
angular momentum operator --see Eq.(\ref{Pm1m1Opt}) below. In turn, this causes the E1M1 transition probability to depend on  
 the spin polarization of the complex, whereas neither the M1M1 nor the E1E1 probabilities do. Note also that stimulated
emission from the state $|\Psi \rangle $ is absent in optical MChD.  All in all, this
implies that $g_{O}^{D/L}$ is proportional to the magnetization of the sample, which is itself proportional to 
the degree of spin-polarization along $\mathbf{B}_{0}$, $\cos{\theta}$, in agreement with experiments.
\begin{figure}[h]
\includegraphics[height=12.0cm,width=12.6cm,clip]{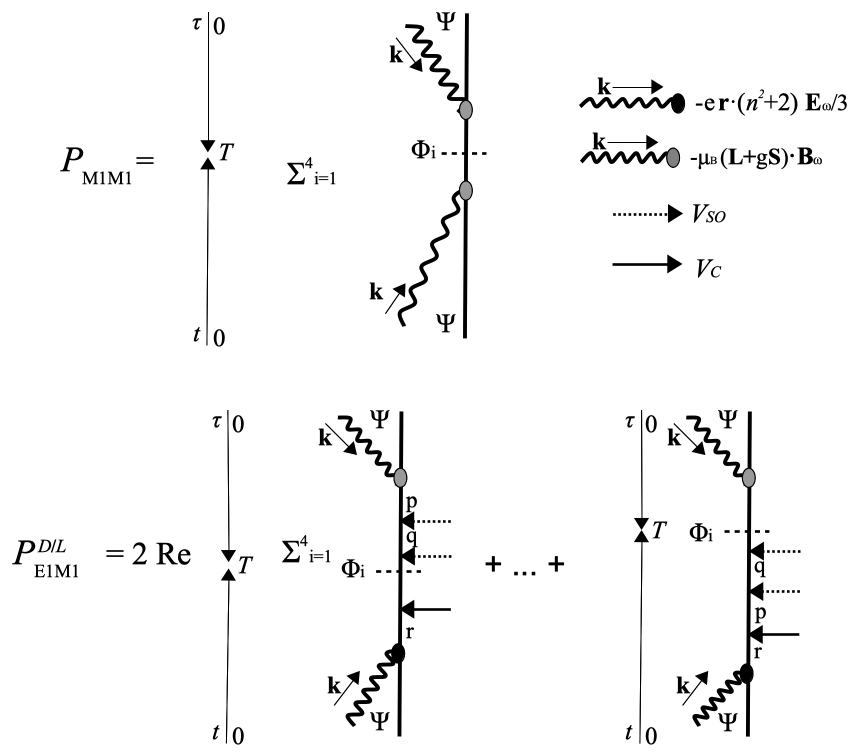}
\caption{Diagrammatic representation of $P_{M1M1}$ and $P_{E1M1}^{D/L}$ for $\omega\approx\Delta_{0}/\hbar$ at
leading order in the perturbative interactions, i.e., at second and up to fifth order, respectively.  
Intermediate atomic states are labeled as $p,q,r,s$.}\label{DiagramsOptics}
\end{figure}
In Fig.\ref{DiagramsOptics}
we depict some of the diagrams which contribute to $P_{M1M1}$ and $%
P_{E1M1}^{D/L}$ in optical MChD. Following a perturbative approach analogous to that
in EPR, for an incident electromagnetic plane wave with $\mathbf{k}\parallel 
\mathbf{B}_{0}$ and assuming $\delta\ll\lambda$, one arrives at%
\begin{align}
P_{M1M1}|_{\omega\approx\Delta_{0}/\hbar}& \simeq \frac{\hbar^{-2}\mu _{B}^{2}|B_{\omega }|^{2}}{4[(\omega -\Delta
_{0}/\hbar)^{2}+\Gamma ^{^{\prime }2}/4]},  \label{Pm1m1Opt} \\
P_{E1M1}^{D/L}|_{\omega\approx\Delta_{0}/\hbar}& \simeq \frac{(\bar{n}^{2}+2)}{3}\frac{C^{D/L}\delta }{2m_{e}\omega _{0}^{3}\tilde{\Delta}}
\frac{\hbar^{-2}\mu _{B}^{2}|B_{\omega }||E_{\omega}|}{4[(\omega -\Delta _{0}/\hbar)^{2}+\Gamma
^{^{\prime }2}/4]}\cos {\theta },  \label{Pe1m1Opt} \\
g_{O}^{D/L}& \lesssim\frac{P_{E1M1}^{D/L,O}}{P_{M1M1}^{O}}\Bigr|_{\omega\approx\Delta_{0}/\hbar}\simeq 
\frac{(\bar{n}^{2}+2)}{3\bar{n}}\frac{c\,C^{D/L}\delta\cos{\theta}}{2\,m_{e}\omega_{0}^{3}\tilde{\Delta}},
\label{gTOpt}
\end{align}%
where $\tilde{\Delta}^{-1}=\Delta_{0}^{-1}+\Delta_{2}^{-1}-3\Delta_{1}^{-1}$, and $\Gamma ^{\prime }$ is the linewidth of optical absorption. As
anticipated, the fact that the magnetic dipole transition in $%
P_{E1M1}^{D/L}$ is dominated by the orbital angular momentum operator 
causes its leading order term to depend on the magnetization $\sim\cos{\theta}$. 
Hence, time-reversal invariance happens to be broken by the spin-polarization of the complex. 

\section{Estimate of $g_{T}^{D/L}$}\label{appD}
In the first place, we work out the relationship between $g^{D/L}_{T}$ and $g^{D/L}_{O}$. Comparing Eq.(\ref{Pe1m1EPR}) with Eq.(\ref{Pe1m1Opt}) at resonance, and taking into account Eqs.(\ref{AgTEPR}) and (\ref{gTOpt}),
we arrive at the following relationships, 
\begin{equation}
\frac{P_{E1M1}^{D/L}|_{\omega =\Omega}}{P_{E1M1}^{D/L}|_{\omega =\Delta
_{0}/\hbar}}\simeq \frac{2\hbar \Omega\tilde{\Delta}\Gamma ^{^{\prime }2} }{\Delta_{0}^{2}\Gamma^{2}}, 
\qquad\frac{g_{T}^{D/L}}{g_{O}^{D/L}}\gtrsim\frac{2\hbar \Omega\tilde{\Delta}}{\Delta_{0}^{2}\cos{\theta}}.\label{comparisonEPROpt}
\end{equation}
Next, considering the experimental data obtained in Ref.\cite{MChD CsCuCl3} for $%
g_{O}^{D/L}$ and applying the relationship in Eq.(\ref{comparisonEPROpt}),
we can estimate a lower bound for $g_{T}^{D/L}$. That is, substituting into
Eq.(\ref{comparisonEPROpt}) the experimental values $g_{O}^{D/L}\approx
0.025 $, $\cos {\theta }\approx 0.4$, for $B_{0}=14$T at a temperature of
4.2 K, we obtain $g_{T}^{D/L}\gtrsim 10^{-4}$.

Alternatively, we can estimate $g_{T}^{D/L}$ using the experimental data of
Ref.\cite{MChD CsCuCl3} for the non-reciprocal absorption coefficient of optical MChD, 
$\alpha_{A}=\alpha(\mathbf{B}_{0}\upharpoonleft \upharpoonright \mathbf{k})-\alpha(\mathbf{B}_{0}\downharpoonleft
\upharpoonright \mathbf{k})$. In order to do so, we first write down $\alpha _{A}$ as
a function of $P_{E1M1}^{D/L,O}$ at resonance, 
\begin{equation}
\alpha _{A}=\frac{4c\mu _{0}\rho \Gamma ^{\prime }\Delta _{0}}{|E_{\omega
}|^{2}}P_{E1M1}^{D/L}|_{\omega =\Delta _{0}/\hbar},
\end{equation}%
where $\rho $ is the molecular density of the CsCuCl$_{3}$ complex (mass
density 3.5g/cm$^{3}$). Substituting the expression for $P_{E1M1}^{D/L,O}(%
\omega =\Delta _{0}/\hbar )$ in the above equation and using Eq.(\ref{AgTEPR}%
) we arrive at the equalities, 
\begin{equation}
C^{D/L}\delta\ =\frac{3\hbar ^{2}m_{e}\omega _{0}^{3}\tilde{\Delta}\Gamma ^{\prime }\alpha _{A}%
}{2(\bar{n}^{2}+2)\rho \mu _{0}\mu _{B}^{2}\Delta_{0}\cos {\theta }},\quad g_{T}^{D/L}\ =
\frac{c\,\hbar^{3}\Gamma'\Omega\tilde{\Delta}\alpha_{A}}{2\Delta_{0}^{3}\mu_{0}\mu_{B}^{2}\rho\cos{\theta}}.
\label{gTEPR2}
\end{equation}%
Substituting the experimental values for all the variables in Eq.(\ref{gTEPR2}),
for $B_{0}=14\,$T at a temperature of $4.2\,$K, with $\Gamma ^{\prime }\approx
0.1$eV and $\bar{n}\approx 1.5$, we obtain $g_{T}^{D/L} \approx 1.5\cdot10^{-2}$, in agreement with our
previous lower bound estimate. 

\section{Further comments on the Hamiltonian model}\label{appE}
Despite the success of our model to derive analytical estimates for the MChA factors, there is still room for improvement. 
In the first place, concerning the chiral Hamiltonian $V_{C}$, it 
was written in terms of the local axis of the octahedral structure, $x$, $y$, $z$, while it should be adapted to 
the crystal axis to account for the helical distribution of the active ions along the 
$c$-axis. In fact, the experimental data on $\alpha_{A}$ taken from the literature to estimate $g_{T}^{D/L}$ consider $\mathbf{B}_{0}$ along the $c$-axis. Also, the harmonic oscillator model, which is considered only distorted in the $n=2,l=2$ level, may not be accurate enough to account for the 
intermediate transitions induced by the chiral potential to levels with $n\neq2$. Hence, 
a more accurate confining potential model, though less generic, can be obtained using a more detailed formulation of the crystal field and 
the JT distortion for the particular case of CsCuCl$_{3}$--see, eg., Ref.\cite{Maaskant}.
Finally, our estimate of the unknown combination $C^{D/L}\delta$ in terms of $\alpha_{A}$ [Eq.(\ref{gTEPR2})], involves 
$\bar{n}$-dependent factors [Eq.(\ref{AgTEPR}], which account for effective incident fields, as well as $\rho$-dependent factors. 
For high densities and $\bar{n}\approx1.5$ those factors are likely to depend on near field terms and spatial correlations when evaluated  at the absorption frequency \cite{MePRA}.

\end{widetext}

\end{document}